\documentclass[preprint,showpacs,preprintnumbers,amsmath,amssymb]{revtex4}

\usepackage{graphicx}
\usepackage{dcolumn}
\usepackage{bm}
\usepackage{latexsym}

\begin{document}

\preprint{}

\title{Double layer formation in the expanding region
of an inductively coupled electronegative plasma}

\author{N. Plihon}
\email{plihon@lptp.polytechnique.fr}
 \author{C.S. Corr}
 \author{P. Chabert}
 \affiliation{Laboratoire de Physique et Technologie des Plasmas,
  UMR 7648 , Ecole Polytechnique, 91128 Palaiseau, France}

\begin{abstract}
Double-layers (DLs) were observed in the expanding region of an
inductively coupled plasma with $\text{Ar}/\text{SF}_6$ gas mixtures. No DL was observed in
pure argon or $\text{SF}_6$ fractions below few percent. They
exist over a wide range of power and pressure although they are only 
stable for a small window of electronegativity (typically between
8\% and 13\% of $\text{SF}_6$ at 1mTorr), becoming unstable at
higher electronegativity. They seem to be formed at the boundary
between the source tube and the diffusion chamber and act as an
internal boundary (the amplitude being roughly 1.5$\frac{kT_e}{e}$)between a high electron density, high electron
temperature, low electronegativity plasma upstream (in the
source), and a low electron density, low electron temperature,
high electronegativity plasma downstream.
\end{abstract}

\maketitle

Double layers (DLs) have been studied over the past decades
theoretically, numerically and experimentally (see \cite{raadu}
and references therein). The biggest part of the literature treats
the case of electropositive plasmas, however, DLs were also found
in electronegative plasmas
\cite{kouznetsovjap99,sheri99,sheridlpop99,merlinopofb90}. More
recently, Charles and co-workers ~\cite{charlesapl03,charlespop04}
have observed a current-free DL in the expanding region of a
helicon wave excited plasma at very low pressures (typically less
than a millitorr). A strongly diverging static magnetic field
seemed to be required in order to reach the conditions for DL
formation. Their system also had an abrupt change in radius at the
boundary between the source and the diffusion chambers, which
could possibly be a source of DL formation \cite{andrews71}.

In this letter we show that a DL can be formed in a system that
has a similar geometry to that of Charles and co-workers, but
without the use of a static diverging magnetic field. However,
this was possible only with a minimum percentage of an
electronegative gas (namely $\text{SF}_6$) added to argon.
Moreover, this electronegative DL was mostly unstable. It was
found to be stable for a very narrow range of $\text{SF}_6$ mixtures.

\begin{figure}
\includegraphics{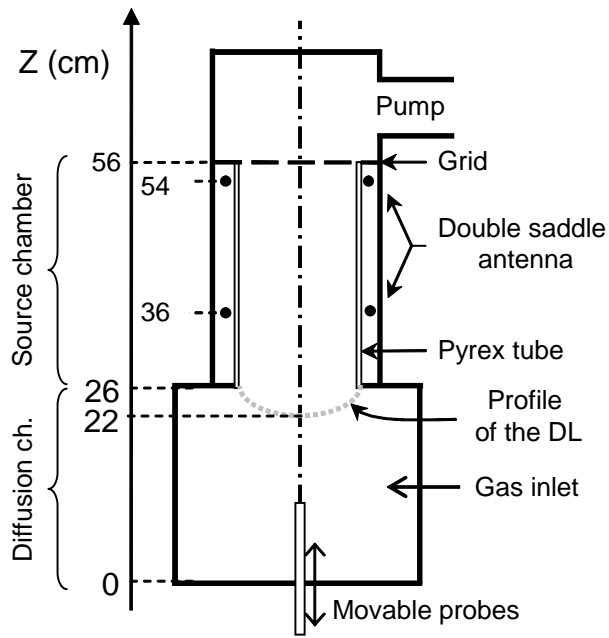}
\caption{\label{plihonfig1}Schematic of the experimental setup.}
\end{figure}

The reactor is shown schematically in figure \ref{plihonfig1}. It was
originally designed to operate in the helicon regime. For
the work presented in this letter, the system was operated without a
static magnetic field, i.e. in the inductive mode. The reactor
consists of a source chamber sitting on top of a 32 cm diameter,
26 cm long aluminum diffusion chamber. Hence, there is an
expanding plasma underneath the source region. The source is a 14
cm diameter, 30 cm long and 0.7 cm thick pyrex cylinder surrounded
by a double saddle field type helicon antenna \cite{bos70}. The
fan-cooled antenna is powered through a close-coupled L-type
matching network by an rf power supply operating at 13.56 MHz and
capable of delivering up to 2 kW forward power. The input power
was recorded as the difference between the forward and reflected
powers. The pyrex cylinder is housed in an aluminum cylinder of 20
cm diameter and 30 cm long. A metal grid attached to the other end
of the source tube confines the plasma from a turbomolecular pump
that routinely achieves base pressures of $10^{-6}$ mbar. The
partial gas pressures of Ar and $\text{SF}_6$ are determined by
controlling the flows.

All measurements reported here were made along the revolution axis
($z$ axis) of the discharge. Two types of electrostatic probes
were used for measurements. The first is a nickel planar probe
with guard ring biased at the same potential as the probe, to
measure the real saturated positive ion current. The diameter of
the collecting area was 4 mm and the diameter of the outer ring
was 8 mm. The second is a passively compensated Langmuir probe (LP)
\cite{can77}, of 0.25 mm diameter and 6 mm long platinum wire tip.
The LP was used to find the plasma potential,
electron densities  and electron temperature from measurements of the probe I(V)
characteristics using a Smartsoft data acquisition system
\cite{hop86}. The electronegativity, $\alpha = n_-/n_e$, and
consequently the ion densities (electro-neutrality $n_+=n_-+n_e$
was assumed), were measured according to the double-probe
technique described in \cite{chabPSST99}. This technique, which
relies on the theory developed in \cite{sheri99}, allows to deduce
$\alpha$ from the ratio of the cylindrical probe current at the
plasma potential to the positive ion saturation current measured
by the planar probe, $R=I(V_p)/I_{\rm sat+}$. The
technique requires an estimation of the ratio of the electron temperature 
to the negative ion
temperature $\gamma = T_e/T_-$ and the
positive ion mass $m_+$, both difficult to measure in the gas
mixture studied here. We chose $\gamma=15$, as is commonly thought
to be a reasonable value in low pressure electronegative
discharges, and $m_+=40$ since (i) $\text{Ar}^+$ may be dominant
since we used small percentages of $\text{SF}_6$ in argon (ii) we
expect a fairly high dissociation degree of $\text{SF}_6$ and
therefore $\text{SF}_x^+$ ions with $x\ll6$ (low mass ions). As a
consequence of these estimations, the absolute values of $\alpha$
should be regarded as indicative. However, we believe that spatial
gradients of $\alpha$, or relative variations with operating
conditions (pressure, power, mixture) are correctly captured by
the technique.

Stable DLs are accessible for pressures from 0.3 to 10 mTorr when carefully adjusting the $\text{SF}_6$ concentration. The minimum power required to obtain a stationary DL increases with increasing pressure, with no upper limit observed (at 1 mTorr, DLs are observed above 200W; at 10 mTorr, above 1400W)

\begin{figure}
\includegraphics{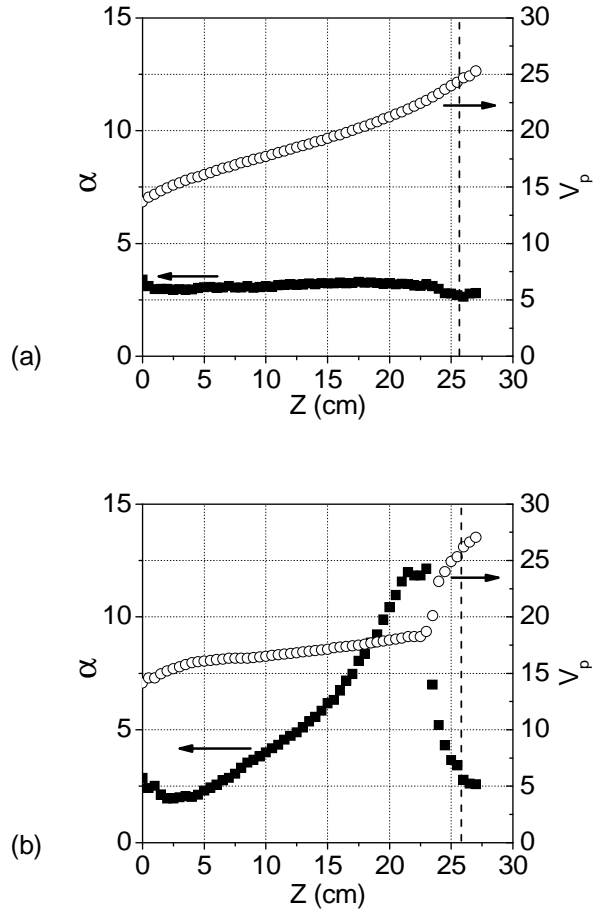}
\caption{\label{plihonfig2} Spatial evolution of the plasma potential,
$V_p$, and the electronegativity, $\alpha = \frac{n_-}{n_e}$, in (a) the no DL case (6\% $\text{SF}_6$ mixture),
and (b) the DL case (9\% $\text{SF}_6$ mixture), at 1 mTorr, 600 W.}
\end{figure}

All results presented here are for a gas pressure of 1 mTorr and an input
power of 600W.
Figure \ref{plihonfig2}a shows the axial evolution of the plasma
potential and the electronegativity for a $\text{SF}_6$
concentration of 6\%. The dashed line represents the position of the
interface between the source and the diffusion chamber. The plasma
potential decreases continuously from the source to the diffusion
chamber, as expected for an expanding plasma which exhibits a
gradient in the electron density, while the electronegativity
remains roughly constant along the axis. There is evidently no DL.
For these conditions of pressure and power, the transition towards the formation
of the DL is observed to occur at about 8\% $\text{SF}_6$
concentration; with no DLs observed in pure argon or for
$\text{SF}_6$ concentrations below 8\%. Above this concentration,
the plasma potential and particles gradients are drastically
changed as shown in figure \ref{plihonfig2}b. The plasma potential
presents a sharp drop at around z = 22cm on the axis, that is about 4 cm below
the interface between the two chambers. The potential difference
between the source chamber and the diffusion chamber seems to be
at least 10V, although the sharp drop seems to be around 5V (which is $1.5\frac{kT_{e}}{e}$ with the downstream electron temperature). This sharp drop is preceded by a strong but
smoother gradient that resembles a pre-sheath. From visual observation, 
it seems that the DL has a
spherical shape that is attached to the boundary between the
source and the diffusion chamber and that expands into the diffusion chamber (refer to the dashed gray line on Figure~\ref{plihonfig1}).

\begin{figure}
\includegraphics{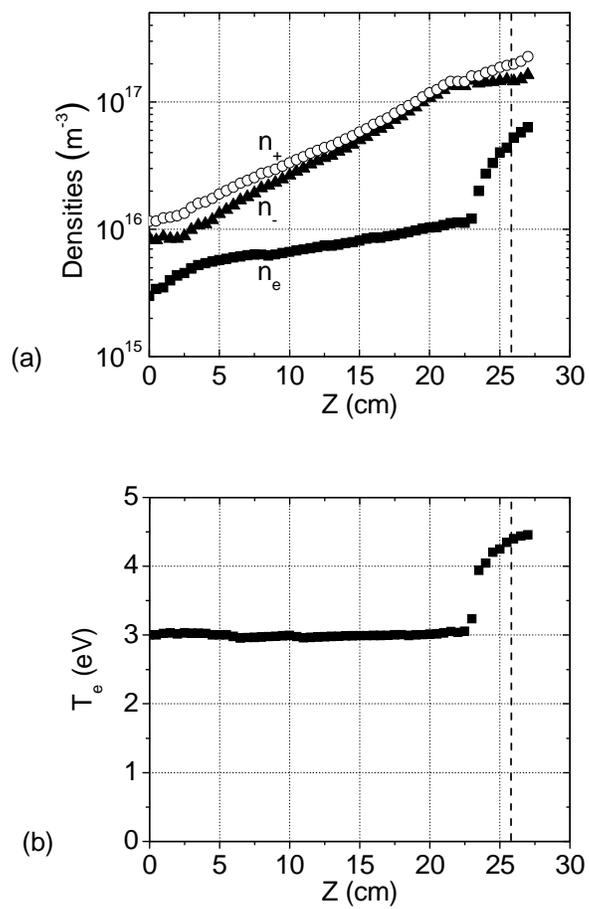}
\caption{\label{plihonfig3} Spatial evolution of the particles densities
and electron temperature for a 9\% $\text{SF}_6$ mixture at 1 mTorr, 600W.}
\end{figure}

The electronegativity is also profoundly affected. It presents a
sharp maximum at the DL position, with a slow decay downstream
(below the DL in the diffusion chamber) and a much faster decay
upstream. The variations of $\alpha$ are directly related to the
change in the electron density, as shown on figure \ref{plihonfig3}a.
The electron density is strongly affected by the sudden drop in
potential, whereas both the positive and negative ion densities
seem to decrease continuously from the source to the diffusion
chamber. The electron temperature changes significantly when
crossing the DL. The DL acts as an internal boundary (or sheath),
which separates two plasmas; a high electron density, high
electron temperature, low electronegativity plasma upstream, and a low
electron density, low electron temperature, high electronegativity
plasma downstream. 

As the $\text{SF}_6$ concentration is increased, the upstream plasma moves further into the diffusion chamber (for a 11\% $\text{SF}_6$ mixture, the spherical shape of the DL being more elongated, the position of the DL on the axis is z = 18cm) and the
plasma potential drop becomes less abrupt, gradually replaced by a
larger region of strong gradient of potential upstream, before
entering the DL itself. The downstream plasma potential remains
mostly constant at about 15 V.

Downstream and upstream of the DL, the electrons remain in Boltzmann equilibrium, with temperatures given by the slope of $\ln(n_e)$ as a function of $V_p$ being 3.2 eV downstream and 4.5 eV upstream, which is very close to the temperatures (from LP processing) given in Figure~\ref{plihonfig3}b. On the
contrary, negative ions are far from Boltzmann equilibrium, and
are present both sides of the DL. Since they cannot cross it from
upstream to downstream (their temperature is much too small), they
must be created downstream, i.e. the attachment rate must be
strong in this region. This may be due to the relatively low
electron temperature, and also to higher neutral gas density
because of colder neutral gas (inductive discharges are known to
produce significant gas heating near the coil). The negative ions
created in the big buffer region downstream from the DL would then be
accelerated toward the source through the DL. Unlike 
attachment, ionization is probably mainly located in the
source region where the the electron temperature is high. Hence,
positive ions are mainly produced in the source region and are
accelerated downstream through the DL. The DL is therefore crossed
by two ion streams in opposite directions.

The origin of the DL formation remains unclear. From our data and
from visual observation, we can postulate that the DL is formed at the boundary between the two
chambers and diffuses in the diffusion chamber, as proposed in earlier work for
electropositive gases \cite{andrews71}. However, this geometric
feature is not sufficient to explain our observations since we did
not observe the DL in pure argon. We can postulate that the
$\text{SF}_6$ addition has two main effect that contributes to the
DL formation. First, the positive ions will more easily reach the
ion sound limitation (a necessary condition to form a DL) since it
is well known that the ion sound speed is lowered in electronegative
plasmas\cite{braith88}. Second, the attachment process is a very efficient loss
term for electrons during the plasma expansion, which makes
steeper $n_e$ gradients and therefore higher potential gradients.
This effect may be compared to the strongly divergent magnetic
field used by Charles and co-workers \cite{charlesapl03}, which
also acts as a loss process for electrons during the expansion.

For the typical conditions considered so far (1 mTorr and 600W),
the DLs were stable from 8\% to 13\% $\text{SF}_6$. Above 13\%,
the DL becomes unstable, and periodic oscillations of the charged
particle densities, plasma potential and electron temperature are
observed. It seems that the unstable regime is characterized by a
periodic formation and propagation of a double layer. This
instability has strong similarities with the downstream
instabilities observed and modeled by Tuszewski and co-workers
\cite{Tuszewskipop03}. However, the association between the
downstream instabilities and a DL was not clearly established by
these authors. This issue will be treated in a separate
publication.

We have observed double-layer formation in the expanding region of
an inductively coupled electronegative plasma. The DL's were not
observed in pure argon. They are stable for a small window of
electronegativity and become unstable at higher electronegativity. They
seem to have a spherical shape and be formed at the boundary
between the source and the diffusion chambers. They act as an
internal boundary between a high electron density, high electron
temperature, low electronegativity plasma upstream, and a low
electron density, low electron temperature, high electronegativity
plasma downstream. They exist in a wide range of pressure and
power and without a strongly divergent magnetic field, which can
be seen as an advantage for space plasma propulsion. However, the
voltage drop in the DL is about three times smaller than the DL's
described by Charles and co-workers \cite{charlesapl03}.

\newpage
\bibliography{095508APL}

\begin{thebibliography}{14}
\expandafter\ifx\csname natexlab\endcsname\relax\def\natexlab#1{#1}\fi
\expandafter\ifx\csname bibnamefont\endcsname\relax
  \def\bibnamefont#1{#1}\fi
\expandafter\ifx\csname bibfnamefont\endcsname\relax
  \def\bibfnamefont#1{#1}\fi
\expandafter\ifx\csname citenamefont\endcsname\relax
  \def\citenamefont#1{#1}\fi
\expandafter\ifx\csname url\endcsname\relax
  \def\url#1{\texttt{#1}}\fi
\expandafter\ifx\csname urlprefix\endcsname\relax\def\urlprefix{URL }\fi
\providecommand{\bibinfo}[2]{#2}
\providecommand{\eprint}[2][]{\url{#2}}

\bibitem[{\citenamefont{Raadu}(1989)}]{raadu}
\bibinfo{author}{\bibfnamefont{M.}~\bibnamefont{Raadu}},
  \bibinfo{journal}{Phys.\ Reports} \textbf{\bibinfo{volume}{178}},
  \bibinfo{pages}{25} (\bibinfo{year}{1989}).

\bibitem[{\citenamefont{Kouznetsov et~al.}(1999)\citenamefont{Kouznetsov,
  Lichtenberg, and Lieberman}}]{kouznetsovjap99}
\bibinfo{author}{\bibfnamefont{I.}~\bibnamefont{Kouznetsov}},
  \bibinfo{author}{\bibfnamefont{A.}~\bibnamefont{Lichtenberg}},
  \bibnamefont{and}
  \bibinfo{author}{\bibfnamefont{M.}~\bibnamefont{Lieberman}},
  \bibinfo{journal}{J.\ Appl.\ Phys.} \textbf{\bibinfo{volume}{86}},
  \bibinfo{pages}{4142} (\bibinfo{year}{1999}).

\bibitem[{\citenamefont{Sheridan
  et~al.}(1999{\natexlab{a}})\citenamefont{Sheridan, Chabert, and
  Boswell}}]{sheri99}
\bibinfo{author}{\bibfnamefont{T.}~\bibnamefont{Sheridan}},
  \bibinfo{author}{\bibfnamefont{P.}~\bibnamefont{Chabert}}, \bibnamefont{and}
  \bibinfo{author}{\bibfnamefont{R.}~\bibnamefont{Boswell}},
  \bibinfo{journal}{Plasma\ Sources\ Sci.\ Technol.}
  \textbf{\bibinfo{volume}{8}}, \bibinfo{pages}{457}
  (\bibinfo{year}{1999}{\natexlab{a}}).

\bibitem[{\citenamefont{Sheridan
  et~al.}(1999{\natexlab{b}})\citenamefont{Sheridan, Braithwaite, and
  Boswell}}]{sheridlpop99}
\bibinfo{author}{\bibfnamefont{T.}~\bibnamefont{Sheridan}},
  \bibinfo{author}{\bibfnamefont{N.}~\bibnamefont{Braithwaite}},
  \bibnamefont{and} \bibinfo{author}{\bibfnamefont{R.}~\bibnamefont{Boswell}},
  \bibinfo{journal}{Phys.\ Plasmas} \textbf{\bibinfo{volume}{6}},
  \bibinfo{pages}{4375} (\bibinfo{year}{1999}{\natexlab{b}}).

\bibitem[{\citenamefont{Merlino and Loomis}(1990)}]{merlinopofb90}
\bibinfo{author}{\bibfnamefont{R.}~\bibnamefont{Merlino}} \bibnamefont{and}
  \bibinfo{author}{\bibfnamefont{J.}~\bibnamefont{Loomis}},
  \bibinfo{journal}{Phys.\ Fluids B} \textbf{\bibinfo{volume}{2}},
  \bibinfo{pages}{2865} (\bibinfo{year}{1990}).

\bibitem[{\citenamefont{Charles and Boswell}(2003)}]{charlesapl03}
\bibinfo{author}{\bibfnamefont{C.}~\bibnamefont{Charles}} \bibnamefont{and}
  \bibinfo{author}{\bibfnamefont{R.}~\bibnamefont{Boswell}},
  \bibinfo{journal}{Appl.\ Phys.\ Lett.} \textbf{\bibinfo{volume}{82}},
  \bibinfo{pages}{1356} (\bibinfo{year}{2003}).

\bibitem[{\citenamefont{Charles and Boswell}(2004)}]{charlespop04}
\bibinfo{author}{\bibfnamefont{C.}~\bibnamefont{Charles}} \bibnamefont{and}
  \bibinfo{author}{\bibfnamefont{R.}~\bibnamefont{Boswell}},
  \bibinfo{journal}{Phys.\ Plasmas} \textbf{\bibinfo{volume}{11}},
  \bibinfo{pages}{1706} (\bibinfo{year}{2004}).

\bibitem[{\citenamefont{Andrews and Allen}(1971)}]{andrews71}
\bibinfo{author}{\bibfnamefont{J.}~\bibnamefont{Andrews}} \bibnamefont{and}
  \bibinfo{author}{\bibfnamefont{J.}~\bibnamefont{Allen}},
  \bibinfo{journal}{Proc.\ Roy.\ Soc.} \textbf{\bibinfo{volume}{A320}},
  \bibinfo{pages}{459} (\bibinfo{year}{1971}).

\bibitem[{\citenamefont{Boswell}(1970)}]{bos70}
\bibinfo{author}{\bibfnamefont{R.}~\bibnamefont{Boswell}},
  \bibinfo{journal}{Phys.\ Lett.} \textbf{\bibinfo{volume}{33A}},
  \bibinfo{pages}{470} (\bibinfo{year}{1970}).

\bibitem[{\citenamefont{Cantin and Gagne}(1977)}]{can77}
\bibinfo{author}{\bibfnamefont{A.}~\bibnamefont{Cantin}} \bibnamefont{and}
  \bibinfo{author}{\bibfnamefont{R.}~\bibnamefont{Gagne}},
  \bibinfo{journal}{Appl.\ Phys.\ Lett.} \textbf{\bibinfo{volume}{30}},
  \bibinfo{pages}{31} (\bibinfo{year}{1977}).

\bibitem[{\citenamefont{Hopkins and Graham}(1986)}]{hop86}
\bibinfo{author}{\bibfnamefont{M.}~\bibnamefont{Hopkins}} \bibnamefont{and}
  \bibinfo{author}{\bibfnamefont{W.}~\bibnamefont{Graham}},
  \bibinfo{journal}{Rev.\ Sci.\ Instrum.} \textbf{\bibinfo{volume}{57}},
  \bibinfo{pages}{2210} (\bibinfo{year}{1986}).

\bibitem[{\citenamefont{Chabert et~al.}(1999)\citenamefont{Chabert, Sheridan,
  Boswell, and Perrin}}]{chabPSST99}
\bibinfo{author}{\bibfnamefont{P.}~\bibnamefont{Chabert}},
  \bibinfo{author}{\bibfnamefont{T.}~\bibnamefont{Sheridan}},
  \bibinfo{author}{\bibfnamefont{R.}~\bibnamefont{Boswell}}, \bibnamefont{and}
  \bibinfo{author}{\bibfnamefont{J.}~\bibnamefont{Perrin}},
  \bibinfo{journal}{Plasma\ Sources\ Sci.\ Technol.}
  \textbf{\bibinfo{volume}{8}}, \bibinfo{pages}{561} (\bibinfo{year}{1999}).

\bibitem[{\citenamefont{Braithwaite and Allen}(1988)}]{braith88}
\bibinfo{author}{\bibfnamefont{N.}~\bibnamefont{Braithwaite}} \bibnamefont{and}
  \bibinfo{author}{\bibfnamefont{J.}~\bibnamefont{Allen}},
  \bibinfo{journal}{J.\ Phys.\ D:\ Appl.\ Phys.} \textbf{\bibinfo{volume}{21}},
  \bibinfo{pages}{1733} (\bibinfo{year}{1988}).

\bibitem[{\citenamefont{Tuszewski and Gary}(2003)}]{Tuszewskipop03}
\bibinfo{author}{\bibfnamefont{M.}~\bibnamefont{Tuszewski}} \bibnamefont{and}
  \bibinfo{author}{\bibfnamefont{S.~P.} \bibnamefont{Gary}},
  \bibinfo{journal}{Phys.\ Plasmas} \textbf{\bibinfo{volume}{10}},
  \bibinfo{pages}{539} (\bibinfo{year}{2003}).

\end{thebibliography}

\newpage
\newpage

\newpage

\newpage

\end{document}